# Ferroelectric Tunneling Junctions for Edge Computing

Erika Covi
*NaMLab gGmbH*
Dresden, Germany,
erika.covi@namlab.com

Quang T. Duong
*NaMLab gGmbH*
Dresden, Germany,
quang.duong@namlab.com

Suzanne Lancaster
*NaMLab gGmbH*
Dresden, Germany,
suzanne.lancaster@namlab.com

Viktor Havel
*NaMLab gGmbH*
Dresden, Germany,
viktor.havel@namlab.com

Jean Coignus
*CEA, LETI*
Grenoble, France
jean.coignus@cea.fr

Justine Barbot
*CEA, LETI,*
*Univ. Genoble Alpes,*
Grenoble, France
justine.barbot@cea.fr

Ole Richter
*BICS, ZIAM & CogniGron,*
*University of Groningen*
Groningen, Netherlands
o.j.richter@rug.nl

Philip Klein
*BICS, ZIAM & CogniGron,*
*University of Groningen*
Groningen, Netherlands
*CITEC Bielefeld University*
Bielefeld, Germany
p.klein@rug.nl

Elisabetta Chicca
*BICS, ZIAM & CogniGron,*
*University of Groningen*
Groningen, Netherlands
e.chicca@rug.nl

Laurent Grenouillet
*CEA, LETI*
Grenoble, France
laurent.grenouillet@cea.fr

Athanasios Dimoulas
*NCSRD*
Athens, Greece
a.dimoulas@inn.demokritos.gr

Thomas Mikolajick
*NaMLab gGmbH & IHM TUD*
Dresden, Germany
thomas.mikolajick@namlab.com

Stefan Slesazeck
*NaMLab gGmbH*
Dresden, Germany
stefan.slesazeck@namlab.com

*Abstract*—Ferroelectric tunneling junctions (FTJ) are considered to be the intrinsically most energy efficient memristors. In this work, specific electrical features of ferroelectric hafnium-zirconium oxide based FTJ devices are investigated. Moreover, the impact on the design of FTJ-based circuits for edge computing applications is discussed by means of two example circuits.

## I. INTRODUCTION

### A. Edge computing

In edge computing applications, data coming from sensors or from human-machine interfaces are directly processed locally. This approach reduces latency and power consumption which would be induced by data transmission into the cloud, and increases system reliability by lowering the dependency on wireless connections. Additionally, it solves data security and privacy issues by keeping data locally stored in the edge devices. Thus, specific requirements must be met by edge computing devices.

In order to prevent the issues arising from the von Neumann bottleneck [1] which restricts the amount of data transfer due to limited memory and data bus bandwidth [2], the concept of physical separation between computing and memory units should be replaced. Instead, memory elements need to be integrated directly into the computation engines. Eventually, neuro-inspired architectures are envisaged where both logic and memory functionality become synergized together in one synaptic or neural device [3].

Another requirement is an ultra-low power consumption, especially if efficient Artificial Intelligence (AI) algorithms are used in mobile devices for tasks that involve pattern recognition, such as computer vision or speech recognition. Typically, AI systems employ artificial neural networks (ANNs) that so far have been realized mainly in software, as many small-scale and mobile devices currently lack the compute power and memory that would be required to operate ANNs for these tasks. Neuromorphic processing systems take inspiration from biology to emulate the power efficient data processing found in neural systems and to overcome the limitations induced by the von Neumann bottleneck. Recently, several neuromorphic processors have been developed [4] - [9]. However, the synaptic connections in most of these chips are made from pure volatile CMOS. Another step towards power-efficient edge computing is the ability for normally-off computing, where the system can be powered on and off frequently in order to process data only when needed or to cope with power interruptions. Hence, internal states of the system should be stored in a non-volatile manner by using non-volatile memory devices within these circuits. It has been widely discussed that memristors combine logic and memory functionality within one single device. In our work we investigate the adoption of ferroelectric memristors for the realization of non-volatile logic circuits for neuromorphic and edge computing.

### B. Ferroelectric memory devices

In ferroelectric hafnium-zirconium oxide (HZO)-based memory devices the polarization state of the ferroelectric layer is used to store information either as two distinct digital values or with multiple intermediate steps, eventually approaching an analog switching characteristic [10]. In general, the polarization state of the ferroelectric layer is switched by the application of an electrical field that exceeds the coercive field of the material, which is typically 1-2 MV/cm. Fast switching of ferroelectric $HfO_2$ films in the ns-regime has already been demonstrated [11]. However, the switching speed is limited by a time-voltage trade-off resulting from the nucleation limited switching (NLS) process of the ferroelectric HZO [12][13].

While write operation is very similar for all ferroelectric devices, the read operation depends strongly on the device concept. Three different flavors of ferroelectric devices are currently highly researched, which are the ferroelectric capacitor (FeCAP), the ferroelectric field effect transistor (FeFET) and the ferroelectric tunneling junction (FTJ) [14]. In order to determine the stored information from the FeCAP typically the switching current is sensed upon polarization reversal of the device, resulting in a destructive readout operation. In contrast, the data stored in the FeFET is sensed non-destructively as a drain-source current, as the polarization state of the ferroelectric gate insulator modulates the threshold voltage of the devices.

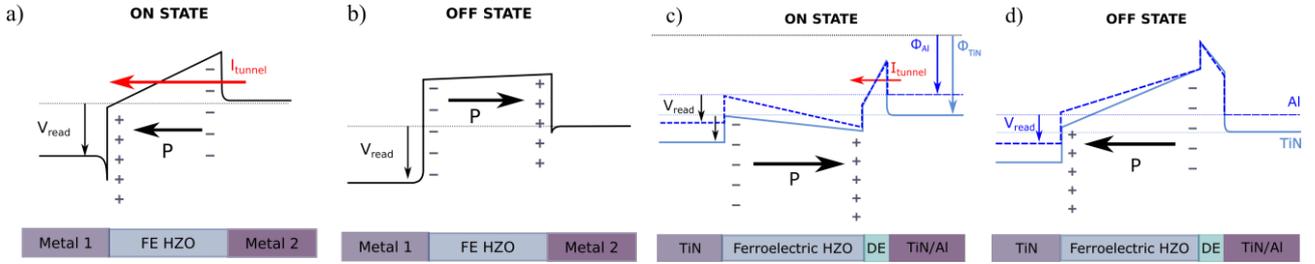

Fig. 1: Schematic view of a band diagram of a single layer FTJ in a) on-state and b) in off-state. c) and d) depict the band diagram of a double layer FTJ in on- and off-state, respectively. The blue arrows indicate the workfunction of the electrode material in direct connection to the tunneling oxide.

In the two-terminal FTJ device, the internal polarization state modifies the electrical resistance of the device stack and can thus be measured as a polarization-dependent current. Typically an electrical field smaller than the coercive field of the ferroelectric layer is applied, in order to avoid a change of the polarization during readout, leading to a non-destructive readout operation. A pinched I-V hysteresis is obtained by reversible voltage sweeps where the internal electrical field exceeds the coercive field of the ferroelectric, thus inducing polarization switching. Hence the FTJ can be considered as a non-linear memristor, as was suggested already for $BaTiO_3$ based FTJ devices [15].

In contrast to most other physical switching mechanisms, the efficiency of the current altering the polarization state of the ferroelectric material is unity. As a result the switching is much more energy efficient compared to other resistive switching mechanisms like valence change, phase change or spin-torque transfer. Moreover, the FTJs feature a low read current as compared to other memristive devices. Therefore, the FTJ is considered as intrinsically the most power-efficient resistance switching device [16]. As such, the FTJ is a very promising non-volatile device that is especially interesting for adoption in circuits targeting at massive parallel processing of data.

## II. FTJ Characteristics

### A. FTJ basic structure

The FTJ device consists of a ferroelectric layer which is sandwiched between two electrodes. Typically, the finite screening length of the electrode materials causes a polarization-dependent band bending which changes the effective barrier height and consequently affects the tunneling resistance, as depicted schematically in Figs. 1 a) and b). The so called tunneling electro-resistance (TER) is measured as the ratio between the resistance of the device in low-resistive state (LRS) and high resistive state (HRS), respectively, with typical values ranging up to 100 [17]. Different electrodes are used to ensure that the opposite effects of both electrodes do not cancel each other out. Diverse physical mechanisms might impact the resistance change. One main obstacle is the need for a very thin ferroelectric layer (< 3 nm) in order to ensure large enough tunneling current densities. However, in polycrystalline ferroelectrics like HZO, the existence of grain boundaries between the ferroelectric grains or defects in such thin film could provide additional leakage current paths that might be independent of the polarization domains and would thus screen the resistance change effect. As a solution a double layer FTJ has been demonstrated, where at least one additional dielectric layer is inserted between one of the electrodes and the ferroelectric layer [18][19][20]. In this configuration the functionality of ferroelectric switching layer and tunneling barrier are separated. The corresponding band diagram is depicted schematically in Figs. 1 c) and d). Thus, thicker ferroelectric films can be used that support a larger TER ratio while maintaining a high current density with typical values in the range of $10^2$-$10^3$ µA/cm².

### B. FTJ programming characteristics

In Fig. 2 a) the typical current voltage characteristic of a double layer FTJ is shown. The device (A) features a 10 nm ferroelectric HZO switching layer and a 2 nm $Al_2O_3$ tunneling layer which are sandwiched between two metallic TiN electrodes. The two curves were obtained by applying a triangular voltage sweep between ± 5.5 V at different frequencies, resulting in slew rates of 11 mV/µs and 0.55 V/µs, respectively. Near the coercive voltage, that is about 2.5 V for the 1 kHz curve, distinct current peaks can be observed, which are the result of the ferroelectric polarization reversal. The switched polarization charge equals the integral of the measured current over sweeping time, depicted as the PV-curve in Fig. 2 b). For this kind of HZO layer the remnant polarization is typically in the range of 20 µC/cm². Moreover, in Fig. 2 a) a constant displacement current can be observed, originating from the non-ferroelectric background permittivity of the FTJ with a value of about 1.3 µF/cm² for the selected device. This self-capacitance is mainly determined by the capacitive series connection of HZO and Al$_2$O$_3$ layers, with permittivities of $k_{HZO}$ = 30 and $k_{Al2O3}$ = 9, respectively. Comparing the measured IV-characteristics at 1 kHz and 50 kHz a much larger current is obtained for higher frequencies. Indeed, the dielectric displacement current scales linearly with the frequency, which of course is not surprising, but which is an important aspect for the circuit design that is often overseen when looking only at the PV hysteresis curve.

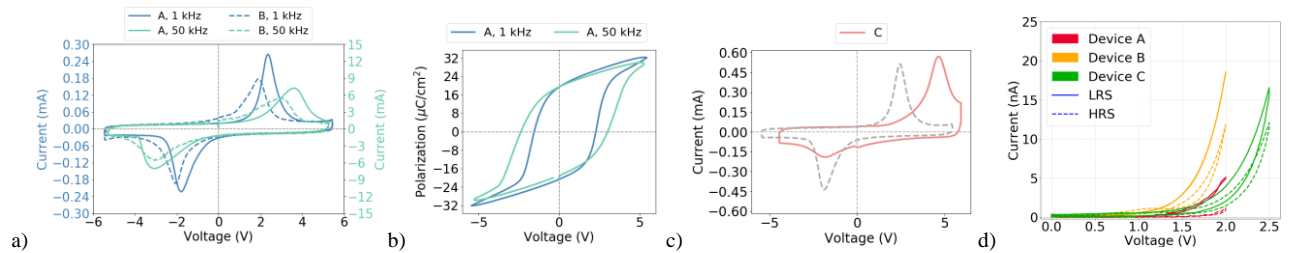

Fig. 2: a) Measured IV-characteristic of TiN/$Al_2O_3$/HZO/TiN FTJ devices for triangular voltage sweeps at 1 kHz and 50 kHz, prepared with two different processes (A – crystallization anneal with TiN electrode, B – crystallization anneal without TiN electrode); b) PV-characteristic as extracted from the data for device A; c) measured IV characteristic of Al/$Al_2O_3$/HZO/TiN FTJ device (labeled C; the grey dashed line shows the TiN/$Al_2O_3$/HZO/TiN device A, for comparison); d) read current of the FTJs shown in a) and c), measured after set/reset pulses of ± 4.5V.

Moreover, from the data in Fig. 2 a) it is clear that at higher frequencies the effective coercive voltage increases up to 4 V. This effect can be explained by the voltage-time trade-off that was reported previously for the NLS polarization reversal in ferroelectric HZO layers [12][13]. When looking from the application perspective, a very steep switching peak is beneficial for digital applications where the FTJ should be switched between two very distinct polarization states (e.g. in non-volatile SRAM (NV-SRAM) cells as discussed in section III B). However, for application as weighting elements in the analog vector-matrix-multiplication based on Ohm's and Kirchhoff's law [21] or as synaptic weighting element [22], a more gradual switching is important because it allows the partial switching of the device in order to attain multiple intermediate resistance states. Hence, not only a proper selection of the programming voltages, but a suitable choice of the programming pulse width is very important and has to be adapted to the specific FTJ device as well as to the voltage constraints that are given from the CMOS technology that interfaces directly to the FTJ.

*C. FTJ reading characteristics*

Fig. 2 d) depicts the DC-IV characteristic of the same FTJ (device A). The two IV-curves are attained after switching the FTJ into LRS or HRS by applying 10 µs long voltages pulses of ± 4.5 V in amplitude, respectively. Note that in this measurement the voltage was swept from 0 V to 2 V (or 2.5 V for device C), which is well below the coercive voltage, in order to prevent polarization switching during the measurement. In this measurement the current is recorded only after the step-wise voltage increase of 50 mV, thus reducing the measured displacement current originating from the non-ferroelectric background capacitance to a minimum. Under these conditions the polarization dependent tunneling current becomes dominant. Compared to the current plateau as visible in Fig. 2 a) at about 20 µA for the 1 kHz sweep, the read current is just in the range of 6 nA and thus more than four orders of magnitude smaller than the displacement current. This ratio increases linearly with the measurement frequency. Obviously, the inherent self-capacitance of the FTJ strongly affects the read operation. This can be avoided in two ways.

In a first method, for an applied read voltage pulse with rising and falling voltage edges, the impact of the displacement current on the read circuit might cancel out. However, as can be seen from the PV-curve in Fig. 2 b), it is obvious that the capacitance of the FTJ depends on the polarization state and applied voltage and thus can be considered as a kind of non-linear mem-capacitor. Hence, depending on the polarization state, the FTJ read operation might be further compromised. In a second method the voltage that is forced to the FTJ during read operation must be kept constant. Consequently, since the FTJ is a two terminal device and thus the programming path and reading path are the same, programming operation and reading operation have to be performed sequentially and cannot be easily performed at the same time. That means, for the application of the FTJ as a synaptic weighting element, special care should be taken in the design and operation of the interfacing circuits.

Another challenge originates from the fact that in typical CMOS circuits there is no way to sense any current without a change in voltage at a certain point within the circuit. In a typical voltage sensing scheme, the FTJ current charges or discharges the capacitance of e.g. a bit line, whereas in a current sensing scheme at least the gate capacitance of a transistor in the reading circuit has to be charged by a certain amount. In the most ideal case where the large area of the FTJ device determines the capacitance of the circuit node to be charged, the charging time $t_{read}$ that is mandatory to attain a certain voltage difference $\Delta V$ depends only on the FTJs current density $J_{FTJ}$ and the FTJs self-capacitance per area $C0_{FTJ}$ according to the equation: $t_{read} = \Delta V \, C0_{FTJ} / J_{FTJ}$. Assuming a minimum voltage difference of $\Delta V$ = 50 mV, a self-capacitance of 1.3 µF/cm² and a current density of 1 pA/µm² as calculated for the device (A) shown in Fig. 2, the reading time $t_{read}$ is as large as 65 ms. Obviously, this reading time is too large for typical memory access. However, this small read current might be the key parameter for applications where a very large number of memory cells are read in parallel, e.g. in the case of NV-SRAM with full parallel restore at power-up, as discussed in Section III.

*D. FTJ device engineering*

As becomes obvious from the discussion above, one of the main optimization targets for the FTJ device is to increase the read current density while keeping the FTJs self-capacitance as low as possible. In the case of the double layer FTJ, this can be achieved by engineering the tunneling oxide as well as the work-function of the electrode that is connected to the tunneling layer. For example, as can be seen from the band structure depicted in Fig. 1 c), the adoption of an Al-electrode (device C) instead of a TiN electrode (device A) could lower the effective tunneling barrier height by about 400 meV (Al: $\varphi_{Al}$ =4.1 eV, TiN: $\varphi_{TiN}$ = 4.5 eV). Indeed, comparing the IV- characteristics of these two FTJ devices in write mode (see Fig. 2 c), obviously a shift of the whole hysteresis to more positive voltages can be seen. Thus, also the effective coercive voltage of the Al/Al$_2$O$_3$/HZO/TiN FTJ is increased and a larger positive read voltage of 2.5 V or 3 V can be applied to the device before polarization reversal sets on, ideally yielding a larger on-current. This current increase was not observed in the experiment, presumably due to an unintentionally decreased effective Al$_2$O$_3$ thickness due to oxide scavenging by the Al electrode, which must be counteracted in future work. Moreover, from the switching peak positions it also becomes obvious that the negative coercive voltage and the corresponding polarization switching peak shift closer towards 0 V. That might cause a data retention issue during storage [23], however, which could be counteracted by changing the work-function of the electrode at the HZO-side as well.

Another optimization goal is the engineering of the width of the switching peak that affects the suitability of the FTJ device for either digital or analog application (as discussed in Section III). In Fig. 2 a) additional polarization switching curves for a TiN/Al$_2$O$_3$/HZO/TiN FTJ device (B) are depicted, the FTJ featuring nominally identical layer thicknesses as device A. However, sample A and B were processed with two different manufacturing processes. In process A the crystallization anneal for the HZO layer was performed after deposition of a TiN capping layer, whereas in process B the same anneal was performed without the capping layer, before deposition of a TiN top electrode of identical thickness. For this process variation, mainly changes in the interface properties and oxidation state between the HZO and Al$_2$O$_3$ layers are expected. Clearly, this modification has a strong effect on the width of the positive polarization switching peak. In summary, a precise control of both thickness of ferroelectric and dielectric layer, as well as of the interface properties and electrode work function is of utmost importance for FTJ optimization targeting at different applications and eventually for well-controlled FTJ functionality.

discharge time constants for synaptic or neuronal integrators. Fig. 3 a) depicts the basic circuit diagram of a differential pair consisting of two FTJs, four access transistors $T_{1n}$, $T_{1p}$, $T_{2n}$, $T_{2p}$ as well as two read transistors $T_3$, $T_4$ that are connected in a similar way as in the 2T1C cell. Moreover, the source terminals of $T_3$ and $T_4$ are tied together to the drain of the bias transistor $T_b$ forming a differential pair. Thus, $I_b$ will always be the sum of $I_1$ and $I_2$ enabling fine control over the normalization behavior. The ratio between $I_1$ and $I_2$ depends on the voltage difference between $n_1$ and $n_2$. Fig. 3 c) shows the simulation results that reveal the functionality of the concept. In this simulation the FTJ model exhibits an about 10 times larger current density than that of device A. Programming of the FTJ devices is performed by adopting the depicted pulsing scheme at BL and PL which ensures that no programming saturation occurs due to the floating nodes $n_1$ and $n_2$. The order of $WL_1$ and $WL_2$ pulses defines which of the FTJs resistance is decreased (write) or increased (erase) to result in the desired weight update. Making use of the NLS kinetics [12], the strength of the FTJs weight update can be adjusted by changing the voltage difference between BL and PL or by the time delay between both signals. Both FTJ devices are always programmed with complementary

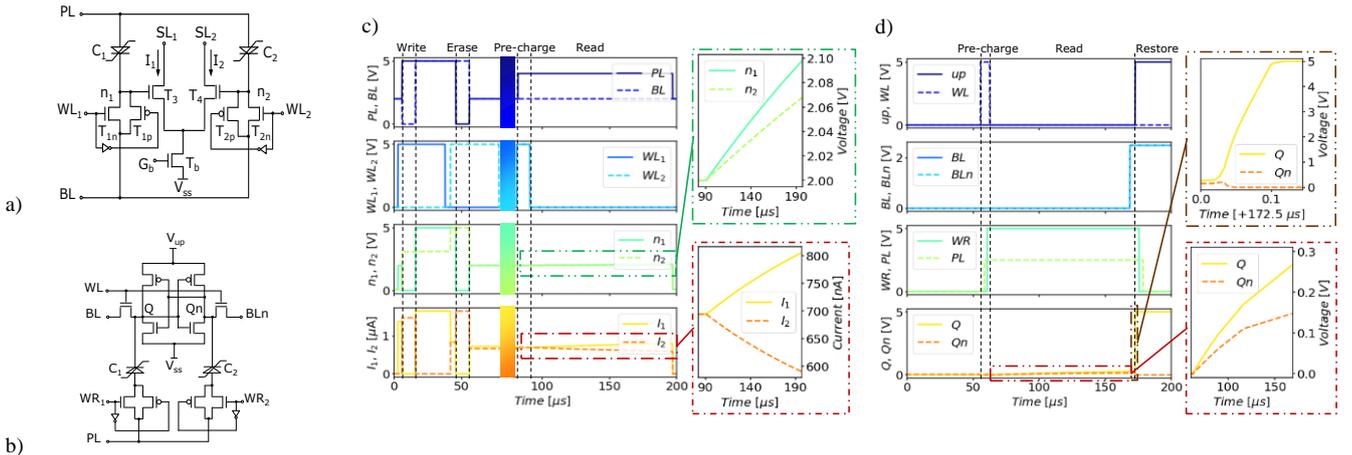

Fig. 3: Circuit schematic of a) differential 2T1C cell and b) NV-SRAM cell; c) simulation of write, pre-charge, and read operations in a differential 2T1C cell. The colored rectangules indicate an arbitrary time span between write and pre-charge operations. Insets: zoom of $n_1$, $n_2$ voltages and $I_1$, $I_2$; d) simulation of read and restore in a NV-SRAM cell as depicted in b). Insets: zoom of Q and $Q_n$ during restore (top) and read (bottom) operations.

## III. FTJ CIRCUITS

As has been reported previously, the small FTJ current can be integrated on the tiny cell-internal capacitance of a 2T1C memory cell featuring one access transistor connected to the FTJ and one read transistor connected to the resulting internal node. The generated voltage signal is then amplified by the read transistor [14]. More complex circuits such as FTJ enhanced differential pairs to be used as normalized synaptic weighting element [24] or NV-SRAM cells [25] are conceived by utilizing the aforementioned concept and are discussed below.

### A. Differential pair for normalized synaptic weighting cells

The normalization behavior of a differential memristor pair enables to mitigate the impact of device mismatch from fabrication and allows fine control over the dynamic range when used as a synaptic weight [24]. In this configuration FTJs can be also used as continuously active, variable weights (i.e. weight $w$ and $1-w$) to allow for fast sensing of the synaptic weight. Indeed, during the read phase, nodes $n_1$ and $n_2$ are floating and their voltage increases with a time constant which depends on the state (i.e. resistance) of the FTJs. The top inset of Fig. 3 c) shows that in 100 µs, $n_1$ and $n_2$ develop a difference of ~30 mV for $C_1$ programmed in LRS and $C_2$ in HRS, which causes $I_1$ to increase and $I_2$ to decrease (Fig. 3 c), bottom inset) resulting in a difference of about 200 nA in the differential pair.

### B. Non-volatile SRAM cell

The NV-SRAM avoids one of the major disadvantages of classical SRAM, namely the continuous power requirement. By extending SRAM cells with FTJ devices we add the functionality to retrieve and store their internal states in parallel on power up and down respectively, while retaining the rapid access times of SRAM not supported by other non-volatile storage types. Proper size matching when adding the FTJ cell in BEOL over the SRAM area together with standard SRAM design techniques would enable almost the same memory density as in classic SRAM arrays. Recently, neuromorphic processors have been using SRAM-based asynchronous content-addressable memory cells (CAM) for storing information about the synaptic connectivity of a network [26]. The adoption of NV-SRAM would allow a frequent power down and power up, increasing the overall energy efficiency of the system.

In Fig. 3b) the circuit diagram of an NV-SRAM cell is shown. Compared to the differential pair of Fig. 3 a) the difference stage is replaced by the cross-coupled inverter pair that is connected to the supply-voltage $V_{up}$. Similar as in the case of the differential pair, in a first phase a differential voltage signal develops on $n_1$ and $n_2$ while $V_{up}$ is tied to 0 V such that all four transistors of the cross-coupled inverter are switched off. After the difference signal is developed, the SRAM cell is activated by switching $V_{up}$ to $V_{dd}$ (in this case 5 V). The initial differential signal is then amplified to the full digital voltage swing (inset Fig. 3 d).

SUMMARY

FTJ devices have been presented as key enablers of energy-efficient, non-volatile edge computing. Due to the very small read currents and comparatively large capacitance of these devices a thorough co-development in terms of device optimization and operation as well as circuit design is required. Based on electrical characterization results from real FTJ devices we demonstrated two examples of such circuit realizations, potentially solving challenges posed by edge computing power constraints.

ACKNOWLEDGMENT

This work was funded by the European Union's Horizon 2020 research and innovation programme under grant agreement No 871737 and in part by the CogniGron research center and the Ubbo Emmius Funds (UG).